\documentclass[conference]{IEEEtran}

\usepackage{amssymb,amsmath,epsfig,graphicx,theorem,threeparttable}
\usepackage{amsfonts}
\usepackage{bm}
\usepackage{cite}

\newtheorem{Theo}{Theorem}
\newtheorem{Lem}{Lemma}

\begin{document}
\IEEEoverridecommandlockouts
\title{Optimal Distortion-Power Tradeoffs in Sensor Networks: Gauss-Markov Random Processes
\thanks{This work was supported by NSF Grants CCR $03$-$11311$, CCF $04$-$47613$ and CCF $05$-$14846$;
and ARL/CTA Grant DAAD $19$-$01$-$2$-$0011$.}}

\author{Nan Liu \qquad Sennur Ulukus \\
\normalsize Department of Electrical and Computer Engineering \\
\normalsize University of Maryland, College Park, MD 20742 \\
\normalsize {\it nkancy@umd.edu} \qquad {\it ulukus@umd.edu} }

\maketitle

\begin{abstract}
We investigate the optimal performance of dense sensor networks by
studying the joint source-channel coding problem. The overall goal
of the sensor network is to take measurements from an underlying
random process, code and transmit those measurement samples to a
collector node in a cooperative multiple access channel with
feedback, and reconstruct the entire random process at the
collector node. We provide lower and upper bounds for the minimum
achievable expected distortion when the underlying random process
is stationary and Gaussian. In the case where the random process
is also Markovian, we evaluate the lower and upper bounds
explicitly and show that they are of the same order for a wide
range of sum power constraints. Thus, for a Gauss-Markov random process,
under these sum power constraints, we determine
 the achievability scheme that is order-optimal, and express the
minimum achievable expected distortion as a function of the sum
power constraint.
\end{abstract}

\section{Introduction}
With the recent advances in the hardware technology, small cheap
nodes with sensing, computing and communication capabilities have
become available. In practical applications, it is possible to
deploy a large number of these nodes to sense the environment.
In this paper, we investigate the optimal performance of a dense
sensor network by studying the joint source-channel coding
problem. The sensor network is composed of $N$ sensors, where $N$
is very large, and a single collector node. The overall goal of
the sensor network is to take measurements from an underlying
random process $S(t)$, $0 \leq t \leq T_0$, code and transmit
those measured samples to a collector node in a cooperative
multiple access channel with feedback, and reconstruct the entire
random process at the collector node. We investigate the minimum
achievable expected
 distortion and the corresponding achievability scheme when the underlying random process is
Gaussian and the communication channel is a cooperative Gaussian
multiple access channel with feedback.

Following the seminal paper of Gupta and Kumar \cite{Gupta:2000},
which showed that multi-hop wireless ad-hoc networks, where users
transmit independent data and utilize single-user coding, decoding
and forwarding techniques, do not scale up, Scaglione and Servetto
\cite{Servetto:2002} investigated the scalability of the sensor
networks. Sensor networks, where the observed data is correlated,
may scale up for two reasons: first, the correlation among the
sampled data increases with the increasing number of nodes and
hence, the amount of information the network needs to carry does
not increase as fast
 as in ad-hoc
wireless networks; and second, correlated data facilitates
cooperation, and may increase the information carrying capacity of
the network. The goal of the sensor network in
\cite{Servetto:2002} was that each sensor reconstructs the data
measured by all of the sensors using sensor broadcasting. In this
paper, we focus on the case where the reconstruction is required
only at the collector node. Also, in this paper, the task is not
the reconstruction of the data the sensors measured, but the
reconstruction of the underlying random process.

Gastpar and Vetterli \cite{Gastpar:sensor2005} studied the case
where the sensors observe a noisy version of a linear combination
of $L$ Gaussian random variables with equal variances, code and
transmit those observations to a collector node, and the collector
node reconstructs the $L$ random variables. In
\cite{Gastpar:sensor2005}, the expected distortion achieved by
applying separation-based approaches was shown to be exponentially
worse than the lower bound on the minimum expected distortion. In
this paper, we study the case where the data of interest at the
collector node is not a finite number of random variables, but a
random process, which, using Karhunen-Loeve expansion, can be
shown to be equivalent to a set of infinitely many random
variables with varying variances. We assume that the sensors are
able to take noiseless samples, but that each sensor observes only
its own sample.
 Our upper bound is also developed by
using a separation-based approach, but it is shown to be of the
same order as the lower bound, for a wide range of power
constraints for a Gauss-Markov random process.

El Gamal \cite{ElGamal:2005} studied the capacity of dense
sensor networks and found that all spatially band-limited Gaussian
processes can be estimated at the collector node, subject to any
non-zero constraint on the mean squared distortion. In this paper,
we study the minimum achievable expected distortion for
space-limited, and thus, not band-limited, random processes, and
we show that the minimum achievable expected distortion decreases
to zero as the number of nodes increases, unless the sum power
constraint is unusually small.

We first provide lower and upper bounds for the minimum achievable
expected distortion for arbitrary stationary Gaussian random
processes. Then, we focus on the case where the Gaussian random
process is also Markovian, evaluate the lower and upper bounds
explicitly, and show that they are of the same order, for a wide
range of power constraints. Thus, for a Gauss-Markov random
process, under a wide range of power constraints, we determine an
order-optimal achievability scheme, and identify the minimum
achievable expected distortion. Our order-optimal achievability
scheme is separation-based. It is well-known \cite
{cover:book},\cite{Cover:1980}
that in multi-user channels with correlated sources, the
source-channel separation principle does not hold in general, and separation-based
achievability schemes may be strictly suboptimal. However, in this
instance, where we have a multi-user channel with correlated
sources, for a wide range of power constraints, we show that a
separation-based achievability scheme is order-optimal, when the
number of nodes goes to infinity.

The results of this paper provide insights for the design of large
sensor networks that aim at reconstructing the underlying random
process at a collector node. Our results provide the order-optimal
scheme for the operation of the sensor nodes, the number of nodes
needed to be deployed and the power constraint needed to be
employed, in order to achieve a certain overall distortion.

Although, we constrain ourselves
to Gauss-Markov processes in this paper, we believe that our
methods can be extended to more general Gaussian random processes.

\section{System Model} \label{systemmodel}
The collector node wishes to reconstruct a random process $S(t)$,
for $0 \leq t \leq T_0$, where $t$ denotes the spatial position;
$S(t)$ is assumed to be Gaussian and stationary with
autocorrelation function $C(\tau)$. The $N$ sensor nodes are
placed at positions $0=t_1 \leq t_2 \leq \cdots \leq t_N=T_0$, and
observe samples $\mathbf{S}_N=(S(t_1),S(t_2),\cdots,S(t_N))$. For
simplicity and to avoid irregular cases, we assume that the
sensors are equally spaced.
The distortion measure is the squared error,
\begin{align}
d(s(t),\hat{s}(t)) = \frac{1}{T_0}\int_{0}^{T_0}
(s(t)-\hat{s}(t))^2 dt
\end{align}

Each sensor node and the collector node, denoted as node 0, is
equipped with one transmit and one receive antenna. At any time
instant, let $X_i$ and $Y_i$ denote the signals transmitted by and
received at, node $i$, and let $h_{ji}$ denote the channel gain
from node $j$ to node $i$. Then, the received signal at node $i$
can be written as,
\begin{align}
Y_i=\sum_{j=0,j \neq i}^N h_{ji} X_j+Z_i, \qquad i=0,1,2,\cdots,N
\end{align}
where $\{Z_i\}_{i=0}^N$ is a vector of $N+1$ independent and
identically distributed, zero-mean, unit-variance Gaussian random
variables. Therefore, the channel model of the network is such
that all nodes hear a linear combination of the signals
transmitted by all other nodes at that time instant. We assume
that $h_{ij}$ is determined by the distance between nodes $i$ and
$j$, denoted as $d_{ij}$, as $ h_{ij}=d_{ij}^{-\alpha/2} $ for
$i,j=0,1,2,\cdots,N$, and $\alpha$ is the path-loss exponent,
which is typically between 2 and 6 \cite{Rappaport:book}. For
simplicity, we assume that the collector node is at an equal
distance away from all of the sensor nodes, i.e., $h_{i0}=h$, for
$i=1,2,\cdots,N$, where $h$ is some constant, independent of $N$.
The results can be generalized straightforwardly to the case where
$h_{i0}$ are non-identical constants.

We assume that all sensors share a sum power constraint of $P(N)$
which is a function of $N$. For the discussion of distortion-power
tradeoffs of the Gauss-Markov processes, we divide $P(N)$ into
five regions.
\begin{itemize}
\item \emph{Very large}: $P(N)$ is larger than $\frac{e^N}{N}$.
\item \emph{Large}: $P(N)$ is between $\frac{e^{N^{1/3}}}{N}$ and
$\frac{e^N}{N}$. \item \emph{Medium}: $P(N)$ is between
$N^{-\frac{1}{1+\frac{1}{\alpha}}}$ and $\frac{e^{N^{1/3}}}{N}$.
\item \emph{Small}: $P(N)$ is between $N^{-1}$ and
$N^{-\frac{1}{1+\frac{1}{\alpha}}}$. \item \emph{Very small}:
$P(N)$ is no larger than $N^{-1}$.
\end{itemize}
The reason why we divide $P(N)$ into these five regions will be
apparent in Sections \ref{seclowerbound} and \ref{secupperbound}.
The two most interesting cases for the sum power constraint are
$P(N)=N P_{\text{ind}}$ where each sensor has its individual power
constraint $P_{\text{ind}}$, and $P(N)=P_{\text{tot}}$ where all
sensors share a constant total power constraint $P_{\text{tot}}$.
Both of these two cases lie in the \emph{medium} sum power
constraint region.
Our goal is to determine the scheme that achieves the minimum
expected distortion $D^N$ at the collector node for a given total
transmit power constraint $P(N)$, and also to determine the rate
at which this distortion goes to zero as a function of the number
of sensor nodes and the power constraint.

In this paper, we seek to understand the behavior of the minimum
achievable expected
 distortion  when the number of sensor nodes is very
large. We introduce the big-O and big-$\Theta$ notations. We say
that $f$ is O($g$), if there exist constants $c$ and $k$, such
that $|f(N)| \leq c|g(N)|$ for all $N>k$; we say that $f$ is
$\Theta(g)$, if there exist constants $c_1$, $c_2$ and $k$ such
that $c_1|g(N)| \leq |f(N)| \leq c_2|g(N)|$ for all $N>k$. All
logarithms are base $e$. Due to space limitations, all proofs are
omitted here and can be found in \cite{Liu_Ulukus:2005}.

\section{The Gauss-Markov process}
A Gauss-Markov process, also known as the Ornstein-Uhlenbeck
process \cite{Uhlenbeck:1930,Wang:1945}, is defined as a random
process that is stationary, Gaussian, Markovian, and continuous in
probability. It is known that
the autocorrelation function of this process is
\cite{Doob:1942,Breiman:book,Karatzas:book}
\begin{align}
C(\tau)=\frac{\sigma^2}{2 \eta} e^{-\eta |\tau|} \label{nine}
\end{align}
The Karhunen-Loeve expansion \cite{Papoulis:book} of the
Gauss-Markov process yields the eigenfunctions
$\{\phi_k(t)\}_{k=0}^\infty$
\begin{align}
\phi_k(t)=b_k \left(\cos \sqrt{\frac{\sigma^2}{\lambda_k}-\eta^2}
t+ \frac{\eta}{\sqrt{\frac{\sigma^2}{\lambda_k}-\eta^2}} \sin
\sqrt{\frac{\sigma^2}{\lambda_k}-\eta^2} t \right)
\end{align}
where $\{\lambda_k\}_{k=0}^\infty$ are the corresponding
eigenvalues and $b_k$ are positive constants chosen such that the
eigenfunctions $\phi_k(t)$ have unit energy. Even though it is not
possible to express $\{\lambda_k\}_{k=0}^\infty$ in closed form,
they can be bounded as
\begin{align}
\lambda_k' \leq \lambda_k \leq \lambda_k'' \label{bound}
\end{align}
where $\{\lambda_k'\}_{k=1}^\infty$ is defined as
\begin{align}
\lambda_k'=\left\{
\begin{array}{ll}
\frac{\sigma^2 T_0^2}{\left(k+\frac{1}{2}\right)^2 \pi^2+\eta^2 T_0^2}, & k \leq K_0 \\
\frac{\sigma^2 T_0^2}{\left(k+1\right)^2 \pi^2}, & k > K_0
\label{definelambdak1}
\end{array}
\right.
\end{align}
with $K_0=\left\lfloor \frac{\eta^2
T_0^2}{\pi^2}-\frac{3}{4}\right\rfloor$ and $\left\lfloor x
\right\rfloor$ is the largest integer smaller than or equal to
$x$; also, $K_0$ may be negative, in which case, the first line in
(\ref{definelambdak1}) should be disregarded.
$\{\lambda_k''\}_{k=1}^\infty$ is defined as
\begin{align}
\lambda_k''=\left\{
\begin{array}{ll}
\frac{\sigma^2}{\eta^2}, & k \leq 1 \\
\frac{\sigma^2T_0^2}{\left(k-1\right)^2 \pi^2}, & k > 1
\end{array}
\right.
\end{align}

Rate-distortion functions are easier to calculate with
$\{\lambda_k'\}_{k=0}^\infty$ and $\{\lambda_k''\}_{k=0}^\infty$,
and the two sequences will be used in place of
$\{\lambda_k\}_{k=0}^\infty$ to develop lower and upper bounds on
the minimum achievable expected distortion.

\section{A Lower Bound on the Minimum Achievable Expected Distortion} \label{seclowerbound}
\subsection{Arbitrary Stationary Gaussian Random Processes}
Let $D^N$ be the minimum achievable expected distortion at the
collector node for a given total transmit power constraint $P(N)$.
In this section, we will develop two lower bounds on $D^N$. We
obtain our first lower bound by assuming that the communication
links from the sensor nodes to the collector node are noise and
interference free. Let $D^{N}_s$ be the MMSE (minimum mean squared
error) when the collector node estimates the underlying random
process by using the exact values of all of the samples taken by
the sensors. Then, it is straightforward to see that,
\begin{align}
D^N \geq D_s^N \label{firstlowerbound}
\end{align}
Since the random process is Gaussian, calculating $D_s^N$ is a
Gaussian MMSE estimation problem. It suffices to consider the
linear MMSE estimator and the resulting expected distortion is
\begin{align}
D_s^N= \frac{1}{T_0}\int_{0}^{T_0} \left(C(0)-\bm{\rho}_N^T(t)
\Sigma_N^{-1} \bm{\rho}_N(t) \right) dt
\end{align}
where
\begin{align}
\bm{\rho}_N(t)=\begin{bmatrix} C(t-t_1) & C(t-t_2) & \cdots &
C(t-t_N)
\end{bmatrix}^T
\end{align}
and
\begin{align}
\Sigma_N & =E[\mathbf{S}_N \mathbf{S}_N^T] \nonumber \\
&= \left[
\begin{array}{cccc}
C(0) & C(t_2-t_1) & \cdots & C(t_N-t_1)\\
C(t_2-t_1) & C(0) & \cdots & C(t_N-t_2)\\
\vdots & \vdots & \vdots & \vdots\\
C(t_N-t_1)& C(t_N-t_2) & \cdots & C(0)
\end{array} \right]
\end{align}

We obtain our second lower bound by assuming
 that all of the sensors know the random process exactly, and,
the sensor network forms an $N$-transmit 1-receive antenna
point-to-point system to transmit the random process to the
collector node. Let $C_u^N$ be the capacity of this point-to-point
system and $D_p(R)$ be the distortion-rate function of the random
process $S(t)$ \cite{Berger:book}. In this point-to-point system,
the
 separation principle holds and feedback does not increase the capacity, and therefore
\begin{align}
D^N \geq D_p(C_u^N)
\end{align}
To evaluate $D_p(C_u^N)$, we first find the rate distortion
function, $R(D)$, of $S(t)$ \cite[Section 4.5]{Berger:book} as,
\begin{align}
R(\theta)= \sum_{k=0}^\infty \max \left(0, \frac{1}{2} \log
\left(\frac{\lambda_k}{\theta} \right) \right) \label{rate}
\end{align}
and
\begin{align}
D(\theta)=T_0^{-1} \sum_{k=0}^\infty \min (\theta,
\lambda_k)\label{distortion}
\end{align}
It can be seen that the function $R(\theta)$ is a strictly decreasing
function of $\theta$ when $\theta \leq \lambda_0$. Hence, in this
region, the inverse function of $R(\theta)$ exists, which we will
call $\theta(R), R \geq 0$. Next, we find $C_u^N$, the capacity of
the $N$-transmit 1-receive antenna point-to-point system
 \cite{Telatar:1999} as,
\begin{align}
C_u^N=\frac{1}{2} \log \left(1+h^2 N P(N) \right) \label{Cupper}
\end{align}
Then, we have
\begin{align}
D_p(C_u^N)=D\left(\theta\left(C_u^N\right)\right) \label{eqn}
\end{align}

By combining the two lower bounds described above, we see that,
for arbitrary stationary Gaussian random processes, a lower bound
on the minimum achievable expected distortion is
\begin{align}
D_l^N=\max \left(D_s^N,D_p(C_u^N)\right) \label{ulukus1}
\end{align}

\subsection{The Gauss-Markov Process}

We note that $D_s^N$ and $D_p(C_u^N)$ in (\ref{ulukus1}) both
depend on the autocorrelation function $C(\tau)$. Unless we put
more structure on $C(\tau)$, it seems difficult to continue with
an exact evaluation. Hence, we constrain ourselves to a special
class of Gaussian random processes, the Gauss-Markov random
processes, whose autocorrelation function is given in (\ref{nine}), in order to continue with our analysis of the distortion.

First, we evaluate $D_s^N$. 
Using (\ref{nine}) and the Markovian property of $S(t)$, it is
straightforward to show that \cite{Liu_Ulukus:2005},
\begin{align}
D_s^N
& = \Theta \left(  N^{-1}\right) \label{DsN}
\end{align}
Hence, for the Gauss-Markov process when the random process is
estimated from its samples, the estimation error decays as
$N^{-1}$.

Next, we evaluate $D_p(C_u^N)$ for the Gauss-Markov process. Let
$D'_p(C_u^N)$ be the distortion obtained from (\ref{eqn}) when
$\{\lambda_k\}_{k=0}^\infty$ in (\ref{rate}) and
(\ref{distortion}) are replaced by $\{\lambda_k'\}_{k=0}^\infty$
which we defined in (\ref{bound}) and (\ref{definelambdak1}).
Then,
\begin{align}
D'_p(C_u^N) \leq D_p(C_u^N) \label{Tunis}
\end{align}
because $\lambda_k' \leq \lambda_k$ for all $k$, and it is more
difficult to estimate a sequence of random variables each with a
larger variance. Since we seek a lower bound on the minimum
achievable expected distortion, the evaluation of $D'_p(C_u^N)$
suffices. Hence, for the rest of this section, we concentrate on
the evaluation of $R(\theta)$ and $D(\theta)$ given in
(\ref{rate}) and (\ref{distortion}), respectively, for
$\{\lambda_k'\}_{k=0}^\infty$.

We will divide our discussion into two separate cases based on the
sum power constraint. For the first case, $P(N)$ is such that
\begin{align}
\lim_{N \rightarrow \infty} \left( NP(N)\right)^{-1}=0
\label{powerconstraint}
\end{align}
is satisfied. This includes all sum power constraint regions
defined in Section \ref{systemmodel} except the \emph{very small}
sum power constraint. The cases where $P(N)=N P_{\text{ind}}$ and
$P(N)=P_{\text{tot}}$ are included in $P(N)$ satisfying
(\ref{powerconstraint}). Note from (\ref{Cupper}) that, in this
case, $C_u^N$ increases monotonically in $N$. Since we are
interested in the number $\theta(C_u^N)$, we consider the region
when $R$ is very large for the function $\theta(R)$.
\begin{Lem} \label{cut1}
For large enough $R$, we have
\begin{align}
\theta(R) \geq \left(\frac{\sigma T_0}{2 \pi R} \right)^2
\label{thetastar}
\end{align}
\end{Lem}
We bound $D(\theta)$ for small enough $\theta$ in the next lemma.
\begin{Lem}
For small enough $\theta$, we have
\begin{align}
D(\theta)
& \geq \frac{ \sigma}{ \pi} \sqrt{\theta} \label{largeN2}
\end{align}
\end{Lem}

We are now ready to calculate the distortion. When $N$ is large
enough, using (\ref{Cupper}), (\ref{thetastar}) and
(\ref{largeN2}), we have
\begin{align}
D'_p(C_u^N)
\geq \Theta \left(\left(\log(NP(N))\right)^{-1}
\right)\label{lower2}
\end{align}
We conclude, based on (\ref{ulukus1}), (\ref{DsN}), (\ref{Tunis}) and
(\ref{lower2}), that when the sum power constraint $P(N)$
satisfies (\ref{powerconstraint}), a lower bound on the minimum
achievable expected distortion is
\begin{align}
D^N \geq \Theta \left(\max \left(N^{-1},
\left(\log(NP(N))\right)^{-1} \right) \right)
\end{align}

For the second case, $P(N)$  is such that (\ref{powerconstraint})
is not satisfied.
 $C_u^N$ is either a constant independent of
 $N$ or goes to zero as $N$ goes to infinity. Examining (\ref{rate}), we see that $\theta(C_u^N)$ is bounded
 below by a constant independent of $N$, and hence, $D'_p\left(C_u^N\right)$ is a constant and does not
 go to zero as $N$ increases.

Therefore, for all possible power constraints $P(N)$, a lower
bound on the minimum achievable expected distortion is
\begin{align}
D^N \geq \Theta \left(\max \left(N^{-1}, \min
\left(\left(\log(NP(N))\right)^{-1}, 1 \right) \right) \right)
\end{align}
which can also be expressed ``order-wise'' as
\begin{align}
\left\{
\begin{array}{ll}
N^{-1} & \text{ if } \lim_{N \rightarrow \infty} \frac{e^N}{NP(N)} =0\\
1 & \text { if } \lim_{N \rightarrow \infty} NP(N) =0 \\
\left(\log(NP(N))\right)^{-1} & \text { otherwise }
\end{array}
\right.  \label{finallowerbound}
\end{align}

The first case in (\ref{finallowerbound}) corresponds to the
\emph{very large} sum power constraint defined in Section
\ref{systemmodel}. This is the scenario where the sum power
constraint grows almost exponentially with the number of nodes.
The transmission power is so large that the communication channels
between the sensors and the collector node are as if they are
perfect, and we are left with the ``unavoidable'' distortion of
$N^{-1}$ which we have in reconstructing the random process from
the ``perfect'' knowledge of its samples. Even though this
provides the best performance among all three cases, it is
impractical since sensor nodes are low energy devices and it is
often difficult, if not impossible, to replenish their batteries.

The second case in (\ref{finallowerbound}) corresponds to the
\emph{very small} sum power constraint defined in Section
\ref{systemmodel}. The transmission power is so low that the
communication channels between the sensors and the collector node
are as if they do not exist. The estimation error is on the order
of 1, which is equivalent to the collector node blindly estimating
$S(t)=0$ for all $t \in [0, T_0]$. Even though the consumed power
$P(N)$ is very low in this case, the performance of the sensor
network is unacceptable; even the lower bound on the minimum
achievable expected distortion does not decrease to zero with the
increasing number of nodes.

Hence, the meaningful sum power constraints for the sensor nodes
should be in the ``otherwise'' case in (\ref{finallowerbound}),
which includes the \emph{large}, \emph{medium} and \emph{small}
sum power constraints defined in Section \ref{systemmodel}. The
corresponding lower bound on the minimum achievable expected
distortion as a function of the power constraint is
\begin{align}
D^N \geq \Theta \left(\left(\log(NP(N))\right)^{-1} \right)
\end{align}
The two practically meaningful cases of $P(N)=N P_{\text{ind}}$ and
$P(N)=P_{\text{tot}}$ are in this ``otherwise'' case. In both of
these cases, the lower bound on the minimum achievable expected
distortion decays to zero at the rate of $\left(\log
N\right)^{-1}$.

\section{An Upper Bound on the Minimum Achievable Expected Distortion} \label{secupperbound}
\subsection{Arbitrary Stationary Gaussian Random Processes}
Any distortion found by using any achievability scheme will serve
as an upper bound for the minimum achievable expected distortion.
We consider the following separation-based achievable scheme:
First, we perform distributed rate-distortion coding at all sensor
nodes using \cite[Theorem 1]{Flynn:1987}. After obtaining the
indices of the rate-distortion codes, we transmit the indices as
independent messages using the antenna sharing method introduced
in \cite{ElGamal:2005}. The distortion obtained using this scheme
will be denoted as $D_u^N$.

We apply \cite[Theorem 1]{Flynn:1987}, generalized to $N$ sensor
nodes in \cite[Theorem 1]{Chen:2004}, to obtain an achievable
rate-distortion point. We will consider the case when all sensor
nodes transmit their data at identical rates, and this rate is
determined by the ratio of the sum rate and $N$. We have the
following theorem.
\begin{Theo}
The following sum rate and distortion are achievable,
\begin{align}
D_a^N(\theta')&  = C(0)-\frac{1}{N-1}\int_{0}^{T_0}
\bm{\rho}_N^T(t) \left(\Sigma_N'+\theta' I\right)^{-1}
\bm{\rho}_N(t)  dt \\
R_a^N(\theta') & =\sum_{k=0}^{N-1} \frac{1}{2} \log
\left(1+\frac{\mu_k^{(N)'}}{\theta'} \right)
\end{align}
where $\Sigma_N'=\frac{T_0}{N-1}\Sigma_N$ and
$\mu_0^{(N)'},\mu_1^{(N)'},\cdots,\mu_{N-1}^{(N)'}$ are the
eigenvalues of $\Sigma_N'$.
\end{Theo}
We further evaluate $D_a^N(\theta')$ in the next lemma.
\begin{Lem} \label{whatnoname}
For all stationary Gaussian random processes whose autocorrelation
functions satisfy the Lipschitz condition in the interval
$[-T_0,T_0]$, and have finite right derivatives at $\tau=0$, we have
\begin{align}
D_a^N(\theta') = O \left(\max \left(N^{-1}, \frac{1}{T_0}
\sum_{k=1}^N \left(\frac{1}{\theta'}+\frac{1}{\mu_k^{(N)'}}
\right)^{-1} \right) \right) \label{expand}
\end{align}
\end{Lem}
Lemma \ref{whatnoname} tells us that the expected distortion
achieved by using the separation-based scheme is upper bounded by
the maximum of two types of distortion. The first distortion is of
size $N^{-1}$ and the size of the second distortion depends on the
achievable rate of the channel through $\theta'$. We define the
second distortion as
\begin{align}
D_b^N(\theta') = \frac{1}{T_0} \sum_{k=1}^N
\left(\frac{1}{\theta'}+\frac{1}{\mu_k^{(N)'}} \right)^{-1}
\end{align}

Now, we determine an achievable rate for the communication channel
from the sensor nodes to the collector node. The channel in its
nature is a multiple access channel with potential cooperation
between the transmitters and feedback from the collector node. The
capacity region for this channel is not known. We get an
achievable sum rate for this channel by using the idea presented
in \cite{ElGamal:2005}. The following theorem is a generalization
of \cite[Theorem 1]{ElGamal:2005} from a constant power constraint
to a more general power constraint.
\begin{Theo} \label{generalelgamal}
When the sum power constraint $P(N)$ and the path-loss exponent
$\alpha$ satisfy
\begin{align}
\lim_{N \rightarrow \infty} \frac{1}{N
P(N)^{1+\frac{1}{\alpha}}}=0 \label{powersad1}
\end{align}
the following rate is achievable
\begin{align}
C_a^N=\beta \log (NP(N))
\end{align}
where $\beta$ is a positive constant
defined as
\begin{align}
\beta =  \frac{1+ \frac{1}{\alpha} \lim_{N \rightarrow \infty}
\frac{ \log P(N)}{\log (NP(N))}} {4 \left(1+\frac{1}{\alpha}
\right)}
\end{align}
otherwise, $C_a^N$ approaches a non-negative constant as $N
\rightarrow \infty$.
\end{Theo}
Theorem \ref{generalelgamal} shows that when the sum power
constraint is \emph{very large}, \emph{large} or
\emph{medium}, as defined in Section
\ref{systemmodel}, the achievable rate increases with $N$.
Otherwise, the achievable rate is either a positive constant or
decreases to zero, which will result in poor estimation
performance at the collector node.

The function $R_a^N\left(\theta'\right)$ is a strictly decreasing
function of $\theta'$, thus, the inverse function exists, which we
will denote as $\theta_a^N(R)$. Hence, to find $D_u^N$, we first
find $ \theta_a^N\left(C_a^N\right)$, and then,
\begin{align}
D_u^N & = D_a^N \left(\theta_a^N \left(C_a^N\right) \right)
\label{sigmaDeqn2}
\end{align}
We will perform this calculation when the underlying random
process is Gauss-Markov.

\subsection{The Gauss-Markov Process}

The autocorrelation function of the Gauss-Markov process given in
(\ref{nine}) satisfies the conditions of Lemma \ref{whatnoname}.
Hence, (\ref{expand}) is valid, and
\begin{align}
D_u^N= O \left(\max \left(N^{-1}, D_b\left(\theta_a^N
\left(C_a^N\right) \right) \right) \right) \label{markovvalid}
\end{align}

It remains to evaluate $D_b^N \left(\theta_a^N \left(C_a^N\right)
\right)$. We first define two sequences $\vartheta_L^N$ and
$\vartheta_U^N$ which satisfy
\begin{align}
\lim_{N \rightarrow \infty} \frac{1}{\vartheta_L^N N^{2/3}}  =0,
\qquad \lim_{N \rightarrow \infty} \vartheta_U^N  = 0
\label{increasing}
\end{align}
\begin{Lem} \label{nothingleft}
For large enough $N$ and $R$ in the interval of
\begin{align}
\left[\frac{8 \sigma T_0}{\pi \sqrt{\vartheta_U^N}}, \frac{\sigma
T_0}{4 \pi \sqrt{\vartheta_L^N}}\right] \label{partialinterval}
\end{align}
we have
\begin{align}
\left(\frac{\sigma T_0}{4 \pi R}\right)^2 \leq \theta_a^N(R) \leq
\left(\frac{8\sigma T_0}{ \pi R}\right)^2 \label{partial}
\end{align}
\end{Lem}
Hence, for all $P(N)$ that satisfy
\begin{align}
 \lim_{N \rightarrow \infty} \frac{NP(N)}{e^{N^{1/3}}} =0 \label{onlycase}
\end{align}
and (\ref{powersad1}), we have $C_a^N$ in the interval of
(\ref{partialinterval}) and our result of (\ref{partial}) is
applicable.

Now we upper bound $D_b^N(\theta')$.
\begin{Lem} \label{Quinn}
For $\theta' \in \left[\vartheta_L^N, \vartheta_U^N \right]$ for
large enough $N$, we may upper bound $D_b^N(\theta')$ as
\begin{align}
D_b^N(\theta') \leq \frac{12 \sigma}{\pi} \sqrt{\theta'}
\label{nuoconclude1}
\end{align}
\end{Lem}
The proofs of Lemma \ref{nothingleft} and \ref{Quinn} use the fact
that $\mu_k^{(N)'}$ converges to $\lambda_k$ when $N$ is large
\cite{Servetto:2002}. Since the convergence of
$\mu_k^{(N)'}$ to $\lambda_k$ is not uniform in $k$, the results of Lemma \ref{nothingleft} and \ref{Quinn} are valid only when $P(N)$ satisfies
(\ref{onlycase}).

Hence, when $P(N)$ is such that (\ref{onlycase}) and
(\ref{powersad1}) are satisfied, using (\ref{partial}),
(\ref{nuoconclude1}) and the fact that when $R$ is in the interval
of (\ref{partialinterval}), $\theta_a^N(R)$ is in
$\left[\vartheta_L^N, \vartheta_U^N \right]$,
 we have
\begin{align}
D_b^N\left(\theta_a^N \left(C_a^N \right) \right)
 \leq \Theta \left( \left( \log (NP(N)) \right)^{-1} \right)
\end{align}
Therefore, from (\ref{markovvalid}), an upper bound on the minimum achievable
expected distortion is
\begin{align}
D_u^N \leq \Theta \left( \left( \log (NP(N)) \right)^{-1} \right)
\label{upperboundrepeat}
\end{align}
This upper bound on the minimum achievable expected distortion
coincides with the lower bound described in the ``otherwise''
case in (\ref{finallowerbound}). However, it should be noted that,
the ``otherwise'' case in (\ref{finallowerbound}) corresponds to
the \emph{large}, \emph{medium} and \emph{small} sum power
constraints defined in Section \ref{systemmodel}, whereas
(\ref{onlycase}) and (\ref{powersad1}) are satisfied only for the
\emph{medium} sum power constraint.

\section{Comparison of Lower and Upper Bounds for Gauss-Markov
Processes}

Now, we compare the upper bound in (\ref{upperboundrepeat}) and
the lower bound in (\ref{finallowerbound}). In the \emph{very
large} and \emph{large} sum power constraint regions, our methods
do not apply, e.g. (\ref{upperboundrepeat}) is not valid, and we have not shown whether the lower and upper
bounds meet. However, in this region $P(N)$ is larger than
$\frac{e^{N^{1/3}}}{N}$, and this region is not of practical
interest.

In the \emph{medium} sum power constraint region, $P(N)$ is in the
wide range of $N^{-\frac{1}{1+\frac{1}{\alpha}}}$ to
$\frac{e^{N^{1/3}}}{N}$, and our lower and upper bounds do meet
and the minimum achievable expected distortion is
\begin{align}
D^N=\Theta \left(\left(\log (NP(N)) \right)^{-1} \right)
\end{align}
The order-optimal achievability scheme is a separation-based
scheme, which uses distributed rate-distortion coding as described
in \cite{Flynn:1987} and optimal single-user channel coding with
antenna sharing method as described in \cite{ElGamal:2005}.
The practically interesting cases of $P(N)=N P_{\text{ind}}$ and
 $P(N)=P_{\text{tot}}$ fall into this region. In both of these
cases, the minimum achievable expected distortion decreases to
zero at the rate of
\begin{align}
\left(\log N \right)^{-1}
\end{align}
Hence, the power constraint $P(N)=P_{\text{tot}}$ performs as well
as $P(N)=N P_{\text{ind}}$ ``order-wise'', and therefore, in
practice we may prefer to choose $P(N)=P_{\text{tot}}$.

In the \emph{small} sum power constraint region where $P(N)$
ranges from $N^{-1}$ to $N^{-\frac{1}{1+\frac{1}{\alpha}}}$, our
lower and upper bounds do not meet. The lower bound decreases to
zero as $ \left(\log N\right)^{-1} $ but the upper bound is a
non-zero constant. The main discrepancy between the lower and
upper bounds comes from the gap between the lower and upper bounds
on the sum capacity, $C_a^N$ and $C_u^N$, for a cooperative
multiple access channel with feedback. This region should be of
practical interest because in this region, the sum power
constraint is quite low, and yet the lower bound on the distortion
is of the same order as any $P(N)$ which increases polynomially
with $N$. Hence, from the results of the lower bound, it seems
that this region potentially has good performance. However, our
separation-based upper bound does not meet the lower bound, and
whether the lower bound can be achieved remains an open problem.

In the \emph{very small} sum power constraint region, $P(N)$ is
less than $N^{-1}$, and our lower and upper bounds meet and the
minimum achievable expected distortion is a constant that does not
decrease to zero with increasing $N$. This case is not of
practical interest because of the unacceptable distortion.

\section{Conclusion}
We investigate the performance of dense sensor networks by
studying the joint source-channel coding problem. We provide lower
and upper bounds for the minimum achievable expected distortion
when the underlying random process is stationary and Gaussian.
When the random process is also Markovian, we evaluate the lower
and upper bounds, and show that they are both of order $\left(\log
(NP(N)) \right)^{-1}$ for a wide range of sum power constraints
ranging from $N^{-\frac{1}{1+\frac{1}{\alpha}}}$ to
$\frac{e^{N^{1/3}}}{N}$. In the most interesting cases when the
sum power grows linearly with $N$ or is a constant, the minimum
achievable expected distortion decreases to zero at the rate of
$\left( \log N\right)^{-1}$. For a Gauss-Markov process, under
these power constraints, we have found that an order-optimal
scheme is a separation-based scheme, that is composed of
distributed rate-distortion coding \cite{Flynn:1987} and antenna
sharing method for cooperative multiple access channels
\cite{ElGamal:2005}. We expect our results to be generalizable to
more general classes of Gaussian random processes.

\bibliographystyle{unsrt}
\bibliography{ref}

\end{document}